\documentclass[aps,prb,reprint,superscriptaddress,amsmath,longbibliography,amssymb]{revtex4-2}
\usepackage{graphicx}
\usepackage{dcolumn}
\usepackage{bm}
\usepackage[colorlinks=true, linkcolor=blue, citecolor=blue, urlcolor=blue]{hyperref}
\usepackage{epstopdf}
\usepackage[capitalize]{cleveref}

\begin{document}
\preprint{APS/123-QED}

\title{Layer-dependent quantum transport in KV$_{2}$Se$_{2}$O-based altermagnetic tunnel junctions}

\author{Yue Zhao}
\affiliation{School of Microelectronics, Nanjing University of Science and Technology, Nanjing, Jiangsu 210094, China}

\author{Bin Xiao}
\affiliation{School of Microelectronics, Nanjing University of Science and Technology, Nanjing, Jiangsu 210094, China}

\author{Jiawei Liu}
\affiliation{School of Microelectronics, Nanjing University of Science and Technology, Nanjing, Jiangsu 210094, China}

\author{Hui Zeng}
\email{zenghui@njust.edu.cn}
\affiliation{School of Microelectronics, Nanjing University of Science and Technology, Nanjing, Jiangsu 210094, China}

\author{Jun Zhao}
\email{zhaojun@njupt.edu.cn}
\affiliation{Jiangsu Provincial Engineering Research Center of Low Dimensional Physics and New Energy, and School of Science, Nanjing University of Posts and Telecommunications, Nanjing, Jiangsu 210023, China}
\altaffiliation{Authors to whom correspondence should be addressed: zenghui@njust.edu.cn and zhaojun@njupt.edu.cn}

\date{\today}

\begin{abstract}
Magnetic tunnel junction (MTJ) is the key component to enable information access and increasing number of MTJs is integrated to develop high-density spintronic devices. However, continuous miniaturization of the conventional MTJs is hindered by stray magnetic fields. Altermagnets, combining the advantages of both ferromagnets and antiferromagnets, provide a promising alternative to fabricate versatile MTJs with exotic properties, such as giant spin splitting, high intrinsic frequency, and absence of stray fields. Inspired by the altermagnetic metal candidate $\mathrm{KV_2Se_2O}$ reported recently, we design an altermagnetic tunnel junction (AMTJ) based on $\mathrm{KV_2Se_2O}/\mathrm{SrTiO_3}/\mathrm{KV_2Se_2O}$. Using density functional theory combined with non-equilibrium Green's function, we investigate the layer-dependent quantum transport properties and the tunneling magnetoresistance (TMR) of such AMTJ device. Our calculated results reveal that the transmission of the AMTJ device exhibits a pronounced oscillation behavior dependent on the number of layers of the $\mathrm{SrTiO_3}$ semiconductor, which is attributed to the interface configuration determined by parity of the layer number. In odd-layer devices, the electron-rich O–Se interface exhibits a smooth effective potential and enables transverse momentum ($k_{\parallel}$) transport channels, leading to enhanced transmission. In contrast, in even-layer devices, the Ti–Se interface presents a steeper effective potential, impeding quantum transport through transverse momentum ($k_{\parallel}$) channels. A giant TMR of $4.6\times10^{7}$\% is predicted to be relized by using a 4-layer $\mathrm{SrTiO_3}$. Our findings not only provide physical understanding relevant to the quantum transport in AMTJs, but also unveil that the barrier interface engineering is a strategy to tune the magnetoelectric performance.
\end{abstract}

\maketitle


\section{INTRODUCTION}
MTJ, consisting of a thin insulator sandwiched by two ferromagnetic (FM) electrodes, is the key building block for high-density spintronic devices\cite{Tsymbal2003,Dieny2020}. Conventional FM electrodes suffer from inherent stray field and low resonance frequency\cite{Jenkins2020}. Further improvement of device integration and operation frequency is a great challenge. Meanwhile, conventional antiferromagnetic (AFM) materials generally can not generate highly spin-polarized currents due to complete spin degeneracy\cite{Zelezny2018,Liu2025}. Recently, the emergence of altermagnets (AM) has sparked both theoretical and experimental explorations to overcome these limitations since the AM possesses advantages of the FM and AFM materials\cite{Libor2022}. The altermagnets maintain zero net magnetization to resist stray field and momentum-dependent spin splitting to realize spin-polarization\cite{Baltz2018,Song2025}. Many bulk AM candidates have shown tremendous advantages to generate spin-polarized current with high efficiency\cite{Jungwirth2026}, including $\text{RuO}_{2}$\cite{Feng2022,Zhang2025}, MnTe\cite{Krempasky2024}, and KV$_{2}$Se$_{2}$O\cite{Jiang2025,Yan2026} with \textit{d}-wave spin-splitting and CrSb with \textit{g}-wave symmetry\cite{Ding2024}. Various 2D AM candidates have been theoretically proposed recently\cite{Han2025,Weijie2025,Che2026,Xu2026}. These unique properties make AMs promising candidates to overcome conventional MTJ limitations\cite{Samanta2025a}. 

\begin{figure}
    \centering
    \includegraphics[width=1\linewidth]{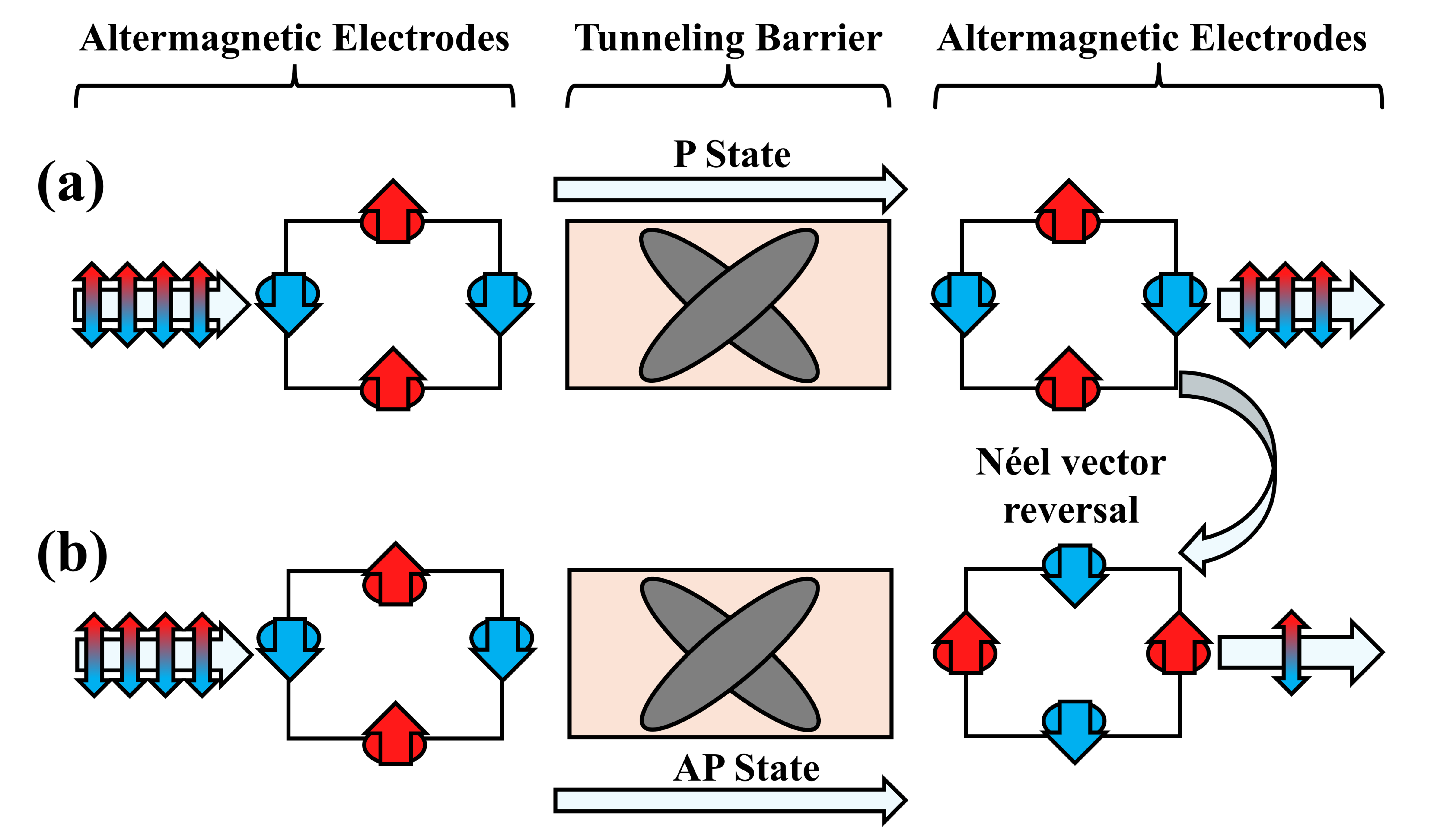}
    \caption{Schematic representation of an altermagnetic tunnel junction device. The device could work under the (a) \textit{P} state or (b) \textit{AP} state, and the two states are switched via manipulating the N\'eel vector of the right electrode. Red and blue circles represent the spin-up and spin-down distributions in the momentum-space (\textit{k}-space) Fermi surface, respectively. The gray ovals in the tunneling barrier region denote the evanescent state. Red and blue solid arrows indicate the spin-polarized currents, and the hollow arrow represents the direction of charge current.}
    \label{FIG-1}
\end{figure}

Consequently, various AMTJs have been proposed in order to achieve a high TMR\cite{Chen2024,Dal2024,Tanaka2025}. The AM metals, such as CrSb and KV$_{2}$Se$_{2}$O, can be used as electrodes to contact with tunneling barrier to fabricate a AMTJ device. As schematically demonstrated in \hyperref[FIG-1]{Fig.~1}, two distinct states, i.e., parallel (\textit{P}) and antiparallel (\textit{AP}) configurations, are defined by parallel and antiparallel for the N\'eel vector of the two electrode, respectively. In the \textit{P} configuration, the momentum-dependent spin splittings in the two electrodes are parallelly aligned, enabling high probability tunneling. In the \textit{AP} configuration, however, the momentum-dependent spin splittings are completely mismatch, hindering quantum tunneling transport\cite{Shao2024,Samanta2024}. As a result, the tunneling transport through the AMTJs can be switched by manipulating the N\'eel vector of the two electrode\cite{Chi2025}. Compared to conventional FM tunnel junction, the AMTJ has a key advantage that momentum-dependent spin splittings in the two electrodes can be fully decoupled, giving rise to almost zero tunneling transport. A giant TMR is anticipated to be obtained in a AMTJ device. For instance, the FM tunnel junction $\mathrm{CoFeB}/\mathrm{MgO}/\mathrm{CoFeB}$ is widely used in commercial magnetic random-access memory (MRAM) currently. The obtained TMR ratio at room temperature typically ranges from 100$\sim$200\%\cite{Zhang2022}. Conversely, the TMR value for most AMTJs could achieve $10^{3}\sim10^{5}$\% according to recent theoretical predictions\cite{Jiang2023,Yang2025,Tanaka2024,zhanglong2025,Zhang2025PRB}, which is much higher than those experimentally achieved in commercial devices. 

Previous study has shown the possibility of improving TMR performance by optimizing the Fermi surface geometry of the two electrode to achieve completely mismatch and match in the \textit{AP} and \textit{P} states, respectively\cite{Lai2025}. More specifically, the momentum-dependent spin channel is generally broad in most three-dimensional bulk altermagnets\cite{Chi2025,Wu2025}. It gives rise to a large region that the two spin transport channels overlap, leading to a prominent leakage current in the \textit{AP} state. The leakage current phenomenon, in turn, can be remarkably suppressed once the momentum-dependent spin channels are completely decoupled\cite{Yang2026}. As a consequence, a perfect "zero-overlap" condition shown in \hyperref[FIG-1]{Fig.~1} is elaborately designed. 

Regarding to practical realization of the AMTJ device shown in \hyperref[FIG-1]{Fig.~1}, the recently fabricated quasi two-dimensional (2D) altermagnetic metals KV$_{2}$Se$_{2}$O\cite{Jiang2025} and Rb$_{1-\delta}$V$_{2}$Te$_{2}$O\cite{Zhang2025NP} provide excellent electrode candidates. The two spin components of the AM KV$_{2}$Se$_{2}$O are almost completely decoupled in \textit{k}-space due to the 2D confinement, as illustrated in \hyperref[FIG-2]{Fig.~2}. Four nodal rings and four nodal arcs are observed in the \textit{k}-dependent Fermi surface, agreeing well with recent experimental observations\cite{Jiang2025,Hu2026}. The spin-resolved transport channels are fully discrete in the \textit{k}-dependent Fermi surface. The quantum transport properties are tuned by manipulating the N\'eel vector of the two electrode, i.e., the "on" and "off" states shown in \hyperref[FIG-1]{Figs.~1(a,b)}. The nonmagnetic (NM) semiconductor SrTiO$_{3}$ is chosen as the barrier layer to construct the KV$_{2}$Se$_{2}$O/SrTiO$_{3}$/KV$_{2}$Se$_{2}$O AMTJ device\cite{Husain2026}. Most importantly, the lattice mismatching between the NM SrTiO$_{3}$ and the AM KV$_{2}$Se$_{2}$O is merely 0.18\%. The interface strain and defect scattering are expected to be substantially suppressed\cite{Goossens2024,Luo2025}. As a consequence, such AMTJ device  proposed by us is experimentally achievable in future. 

In this work, we study the quantum transport properties of the KV$_{2}$Se$_{2}$O/SrTiO$_{3}$/KV$_{2}$Se$_{2}$O AMTJ by using first-principles quantum transport calculations\cite{Taylor2001,Brandbyge2002}. Different numbers of the SrTiO$_{3}$ layer are considered. Our results show that the device's TMR manifests a significant layer-dependent oscillation. This oscillation is attributed to the pronounced charge redistribution resulting from the symmetry and asymmetry of the interface structure. The TMR value of $4.6\times10^{7}$\% is realized by using 4-layer SrTiO$_{3}$ as a barrier. This work provides the physical understanding of the layer-dependent transport in AMTJs, and illustrates that the "barrier-layer engineering" as a route to manipulate quantum tunneling transport to obtain a high-performance TMR for AMTJ based spintronic devices.

\section{METHODS}
The atomic structure and electronic structure calculations are performed by Vienna \textit{ab initio} simulation package (\textsc{VASP})\cite{Kresse1993,Kresse1996}. The calculations are based on density functional theory (DFT), employing the projected augmented wave (PAW) method\cite{Kresse1999,Blochl1994} to describe the interactions between the ion cores and the valence electrons. The exchange-correlation potential is treated within generalized gradient approximation (GGA) using the Perdew-Burke-Ernzerhof (PBE) functional\cite{Perdew1996}. A plane-wave cutoff energy of 450 eV is used. The convergence criteria for energy and force are set to $10^{-6}$ eV and 0.01 eV/{\AA}, respectively. The Brillouin zone integration is sampled by Monkhorst-Pack scheme\cite{Monkhorst1976}. For the structural optimization of the bulk KV$_{2}$Se$_{2}$O and SrTiO$_{3}$, the \textit{k}-mesh is set to $14\times14\times7$ and $11\times11\times11$, respectively. To account for the strong correlation effects of transition metal ions, the GGA+$U$ method is employed\cite{Anisimov1991,Dudarev1998}. Based on the experimental measurement\cite{Jiang2025} and previous report\cite{Yang2026}, the Hubbard $U=0$ for the V-$3d$ orbitals could yield excellent agreement of spin-solved band structure of the AM KV$_{2}$Se$_{2}$O. Hence, the Hubbard $U$ of 6 eV is adopted for the Ti-$3d$ orbitals of the NM SrTiO$_{3}$, which is consistent with previous report shown in Ref. \cite{Chen2020}. The calculated results given by \textsc{VASP} and QuantumATK agree well, as illustrated in Fig. S1 within the Supplemental Material.

\begin{figure}
    \centering
    \includegraphics[width=1\linewidth]{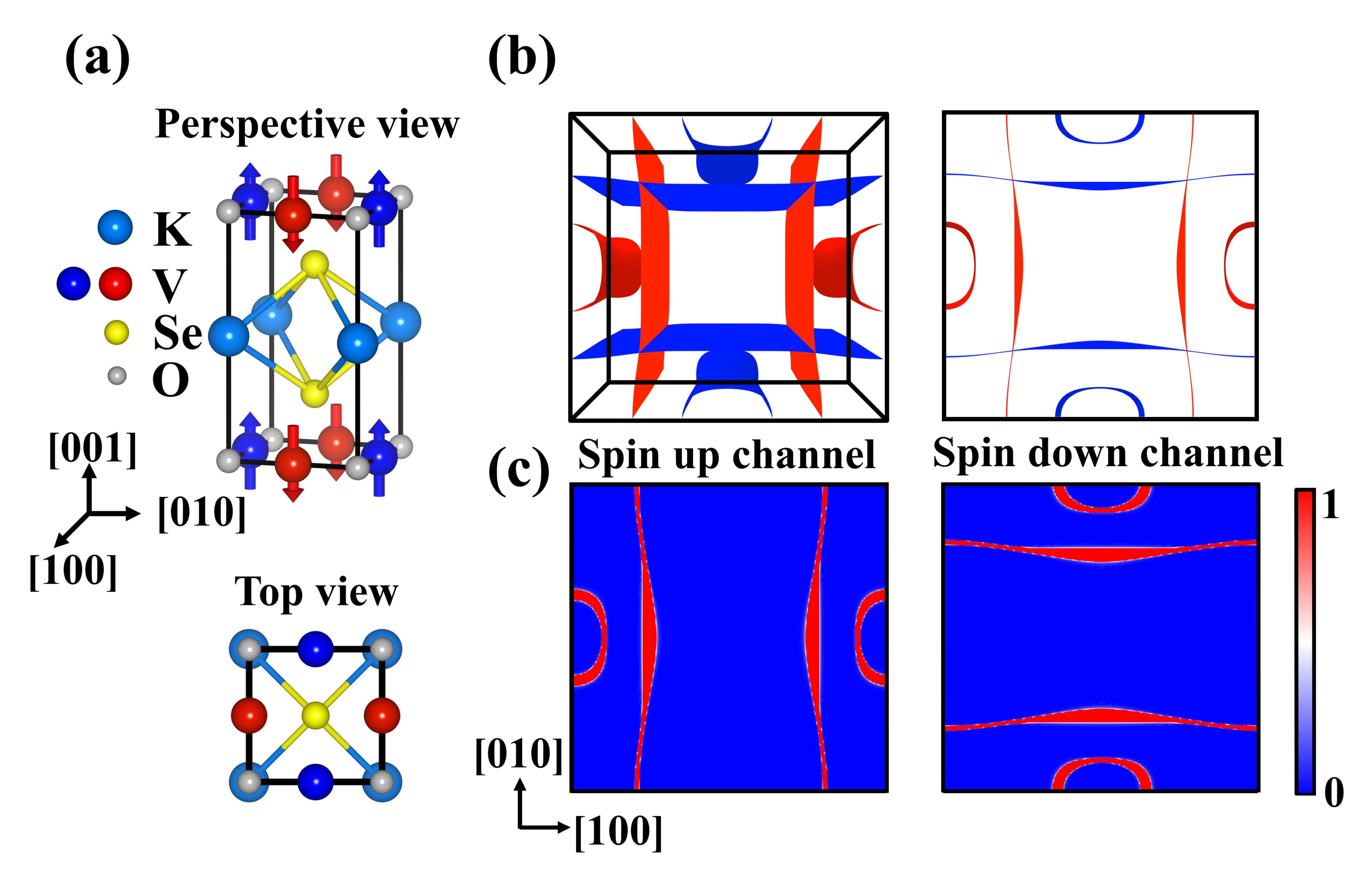}
    \caption{The atomic structure and spin-polarized Fermi surfaces of the $\text{KV}_2\text{Se}_2\text{O}$ crystal. (a) Side and top views of the nanostructure. Two V atoms with opposite magnetic moments are denoted by red and blue spheres. (b) Calculated Fermi surfaces in the $k_{x}-k_{y}$ plane, where red and blue curves represent the spin-up and spin-down components, respectively. The perspective (top) view is shown in the left (right) panel. (c) Spin-resolved transport of the $\text{KV}_2\text{Se}_2\text{O}$ with spin-up (spin-down) component shown in the left (right) panel.}
    \label{FIG-2}    
    \vspace{-3mm}
\end{figure}

The atomic structure and Fermi surface of the metallic $d$-wave AM KV$_{2}$Se$_{2}$O are shown in \hyperref[FIG-2]{Figs.~2(a,b)}. It crystallizes into a tetragonal lattice with space group $P4/mmm$, and it has out-of-plane magnetization axis\cite{Xu2025,Yan2026}. The calculated lattice constant is $a = 3.952$ {\AA}, which is in good agreement with previous studies\cite{Jiang2025,Xu2025}. The two magnetic V atoms carry opposite spin magnetic moments, leading to zero net magnetization. This specific magnetic ordering breaks time-reversal ($\mathcal{T}$) and spatial inversion ($\mathcal{P}$) symmetries. However, the $[C_{2}\parallel C_{4z}]$ and $[C_{2}\parallel M_{1\bar{1}0}]$ symmetries are  preserved, where the operation on the left (right) only acts on the spin (real) space. This symmetry character ensures \textit{k}-dependent spin-splitting, as shown in Fig.~S2. In addition, the calculated spin-resolved Fermi surfaces manifest strong anisotropy. Moreover, the opposite spin transport channels are completely separated in the \textit{k}-space, as evidenced by \hyperref[FIG-2]{Fig.~2(c)}.

The spin transport properties are investigated by using DFT combined with the non-equilibrium Green's function (NEGF)\cite{Taylor2001,Brandbyge2002} formalism, which is implemented in the QuantumWise Atomistix Toolkit (ATK)\cite{Smidstrup2020}. The AMTJ device is modeled by two-probe system composed of left and right KV$_{2}$Se$_{2}$O electrodes and a central SrTiO$_{3}$ scattering region\cite{Fang2024}. In the transport simulation, a mesh cutoff of 150 Hartree is applied, and the electronic temperature is 300 K. The nonrelativistic SG15 pseudopotentials is employed\cite{Hamann2013}. Regarding the Brillouin zone sampling, a \textit{k}-mesh of $14\times14\times201$ is utilized for the bulk electrodes to accurately describe their electronic states, and a $14\times14\times1$ \textit{k}-mesh is employed for the self-consistent calculations of the device region. To obtain accurate $\mathbf{k}_{\parallel}$-resolved transmission spectra, the \textit{k}-mesh is improved to $150\times150$ along the $k_{x}-k_{y}$ direction. Based on the calculated transmission spectra, the TMR ratio is evaluated using the standard definition: $TMR = \frac{T_{P} - T_{AP}}{T_{AP}} \times 100\%$, where $T_{P}$ and $T_{AP}$ denote the total transmission coefficients at the Fermi level ($E_{F}$) for the \textit{P} and \textit{AP} magnetic configurations, respectively.

\section{RESULTS AND DISCUSSION}
\begin{figure}
    \centering
    \includegraphics[width=1\linewidth]{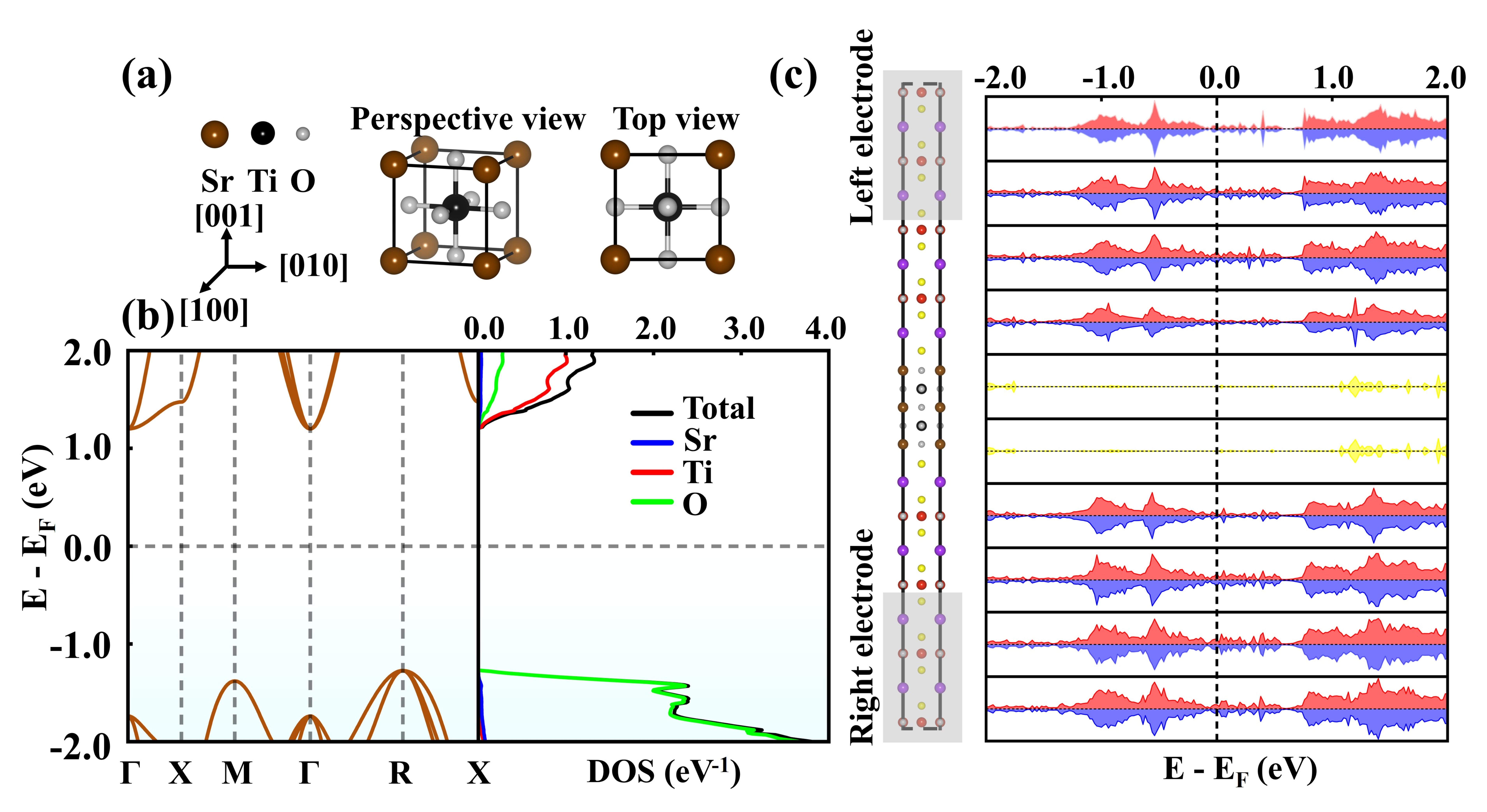}
    \caption{The atomic structure and electronic properties of the $\mathrm{KV_2Se_2O/SrTiO_3/KV_2Se_2O}$ AMTJ. (a) Side and top views of the $\mathrm{SrTiO_{3}}$ crystal. (b) Calculated electronic band structure and projected density of states (PDOS) of the $\mathrm{SrTiO_3}$ barrier. (c) The AMTJ device and the corresponding layer-resolved local density of states (LDOS), where the spin-up and spin-down components are represented by red and blue curves, respectively.}
    \label{FIG-3}
\end{figure}

The optimized lattice constant of the bulk $\mathrm{SrTiO_3}$ is $a = 3.945$ \AA\cite{Liu_2025} and it possess a wide band gap, as shown in \hyperref[FIG-3]{Fig.~3(a,b)}. Moreover, an almost perfect lattice matching between the electrode and the barrier region remarkably suppresses carrier scattering induced by interfacial disorder\cite{Liu2025MH,Yang2025}. The layer-resolved LDOS along the transport direction suggests that the $\mathrm{KV_2Se_2O}$ electrodes exhibit strongly spin-polarized metallic states near the $E_{F}$, as illustrated in\hyperref[FIG-3]{Fig.~3(c)}. The density of states of the central region is suppressed and almost decays to zero. It unveils that the $\mathrm{SrTiO_3}$ layer acts as a high-quality insulating barrier in quantum tunneling phenomenon, hindering electronic state coupling between the two electrodes\cite{Zhao2021,Sun2025,Jiang2023}. 

\begin{figure}
    \centering
    \includegraphics[width=1\linewidth]{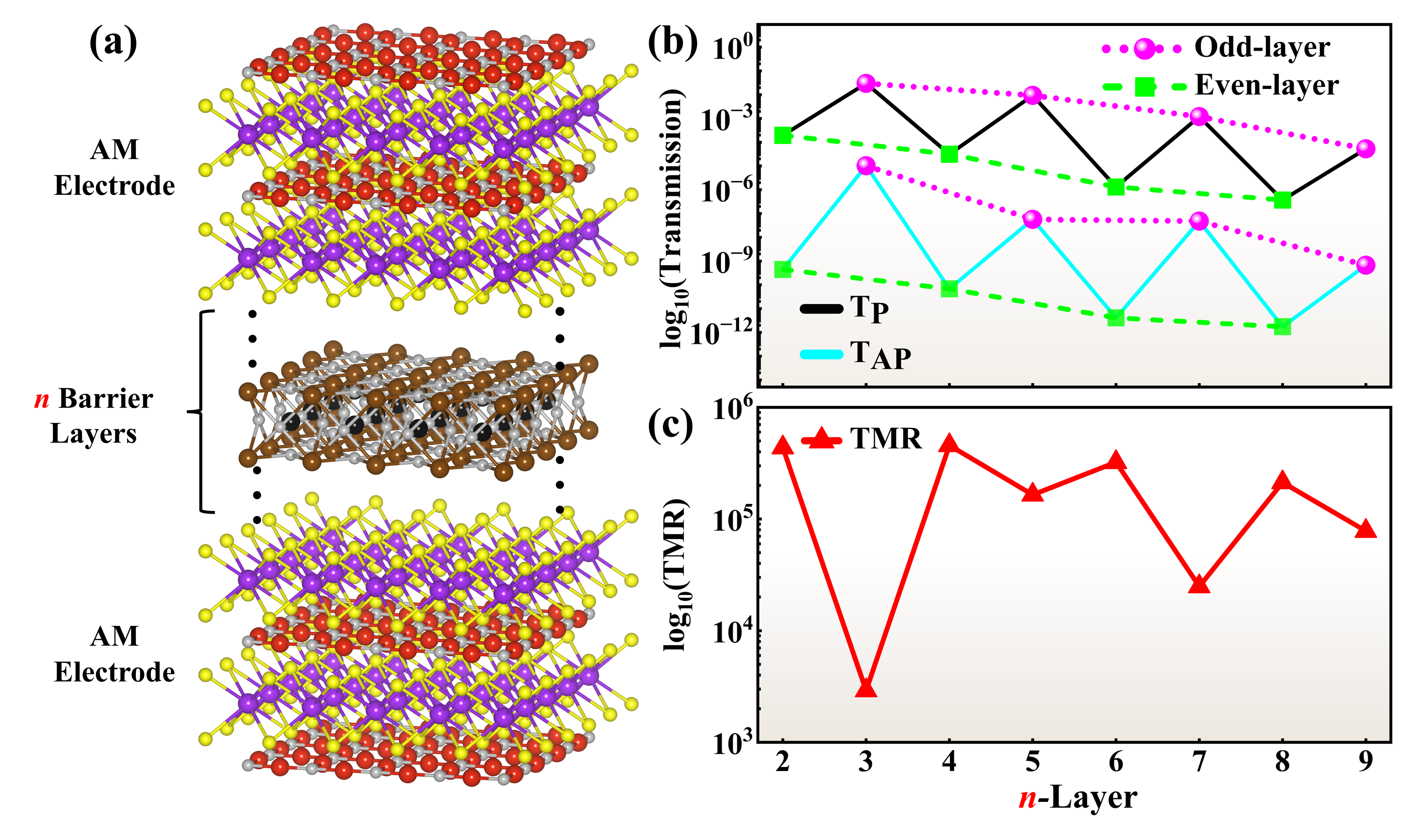}
    \caption{Device structure and layer-dependent spin-transport properties of the $\mathrm{KV_2Se_2O/SrTiO_3/KV_2Se_2O}$ AMTJs. (a) Schematic illustration of the device with $n$-layer SrTiO$_{3}$. (b) Transmission coefficient as a function of the $n$-layer, where the black and cyan solid lines denote $T_{P}$ and $T_{AP}$ obtained at the \textit{P} and \textit{AP} states, respectively. The magenta dotted lines and green dashed lines manifest exponential decay of transmission for the odd-layer and even-layer AMTJs, respectively. (c) Calculated TMR ratio at the $E_{F}$ as a function of the $n$-layer.}
    \label{FIG-4}
\end{figure}

We further investigate the spin transport properties of the $n$-layer AMTJ and examine the layer dependency. As illustrated in \hyperref[FIG-4]{Fig.~4(a)}, the AMTJs with different widths are constructed by varying the numbers of the $\mathrm{SrTiO_3}$ layer ($n = 2, 3, \ldots, 9$). With the increasing of the layer number, the $T_{\mathrm{P}}$ (black solid line) and $T_{\mathrm{AP}}$ (cyan solid line) at the $E_{F}$ exhibit a prominent oscillatory behavior, as shown in \hyperref[FIG-4]{Fig.~4(b)}. The transmission coefficient of the odd-layer AMTJs (magenta solid spheres) systemically larger than those of the even-layer AMTJs (green solid spheres), demonstrating a pronounced odd--even layer dependency. Moreover, the quantum transports in both odd-layer (magenta dotted line) and even-layer (green dashed line) AMTJs are found to exponentially decay. This is in agreement with quantum tunneling. The calculated TMR shown in \hyperref[FIG-4]{Fig.~4(c)} reveals that the even-layer AMTJs exhibit exceptionally superior TMR performance. Specifically, the TMR value of each even-layer AMTJs generally $1\sim2$ orders higher than that of the adjacent odd-layer AMTJs. This is attributed to the significantly larger $T_{AP}$ in the odd-layer AMTJs leading to substantial reduction of the TMR magnitude, because the TMR can be roughly estimated by the ratio of $T_{P}/T_{AP}$. The TMR ratio reaches a maximum of $4.6\times10^{7}$\% when the AMTJ barrier thickness is optimized to 4 layer. 

\begin{figure}
     \centering
     \includegraphics[width=1\linewidth]{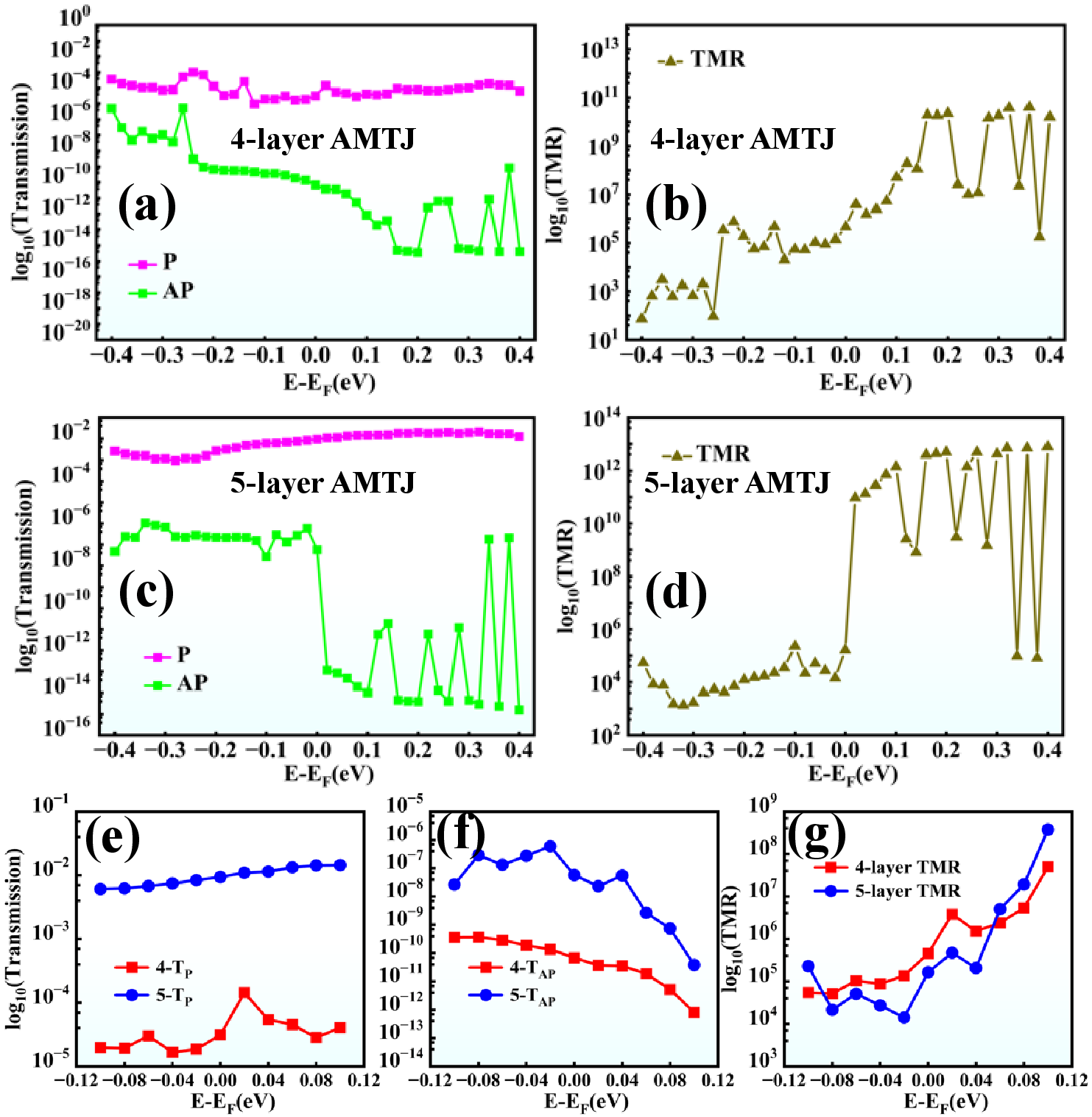}
     \caption{Transport properties of the (a, b) 4-layer and (c, d) 5-layer $\mathrm{KV_{2}Se_{2}O/SrTiO_{3}/KV_{2}Se_{2}O}$ AMTJs. (a) Transmission coefficients under P (magenta) and AP (green) states, and (b) the TMR ratio as a function of energy for the 4-layer AMTJ. (c) and (d) are the corresponding results for the 5-layer AMTJ. Comparison of transmission coefficients at the (e) \textit{P} state and (f) \textit{AP} state, and (g) the calculated TMR for the 4-layer and 5-layer AMTJs.}
     \label{FIG-5}
 \end{figure}

To examine the physical origin of the odd--even layer dependency, we choose the 4-layer and 5-layer AMTJs as an example to compare their differences. The transmission coefficients and TMR of the 4-layer (5-layer) AMTJ are shown in \hyperref[FIG-5]{Figs.~5(a-d)}. For both AMTJs, $T_{\mathrm{P}}$ is higher than $T_{\mathrm{AP}}$. Since the transport is suppressed under the \textit{AP} state, it ensures a stable and giant TMR effect for both AMTJs\cite{Liu2026}. 

Compared \hyperref[FIG-5]{Fig.~5(a)} with \hyperref[FIG-5]{Fig.~5(c)}, the magnitude of the transmission coefficients ($T_{P}$ and $T_{AP}$) of the 4-layer and 5-layer AMTJs are different by several orders. We further extract the transport performance ranging from -0.1 to 0.1 eV, as shown in \hyperref[FIG-5]{Figs.~5(e)-5(f)}. In the vicinity of $ E_{F}$, the $T_{P}$ of the 4-layer AMTJ is merely $4.50\times10^{-5}$, while the corresponding value of the 5-layer AMTJ reaches $1.25\times10^{-2}$. In the \textit{AP} state, the $T_{AP}$ of the 4-layer AMTJ is further reduced to $1.10\times10^{-12}$, and the corresponding value of the 5-layer AMTJ is $2.50\times10^{-8}$. The calculated TMR of the 4-layer AMTJ superiors to that of the 5-layer AMTJ device, as illustrated in \hyperref[FIG-5]{Fig.~5(g)}.

\begin{figure}
    \centering
    \includegraphics[width=1\linewidth]{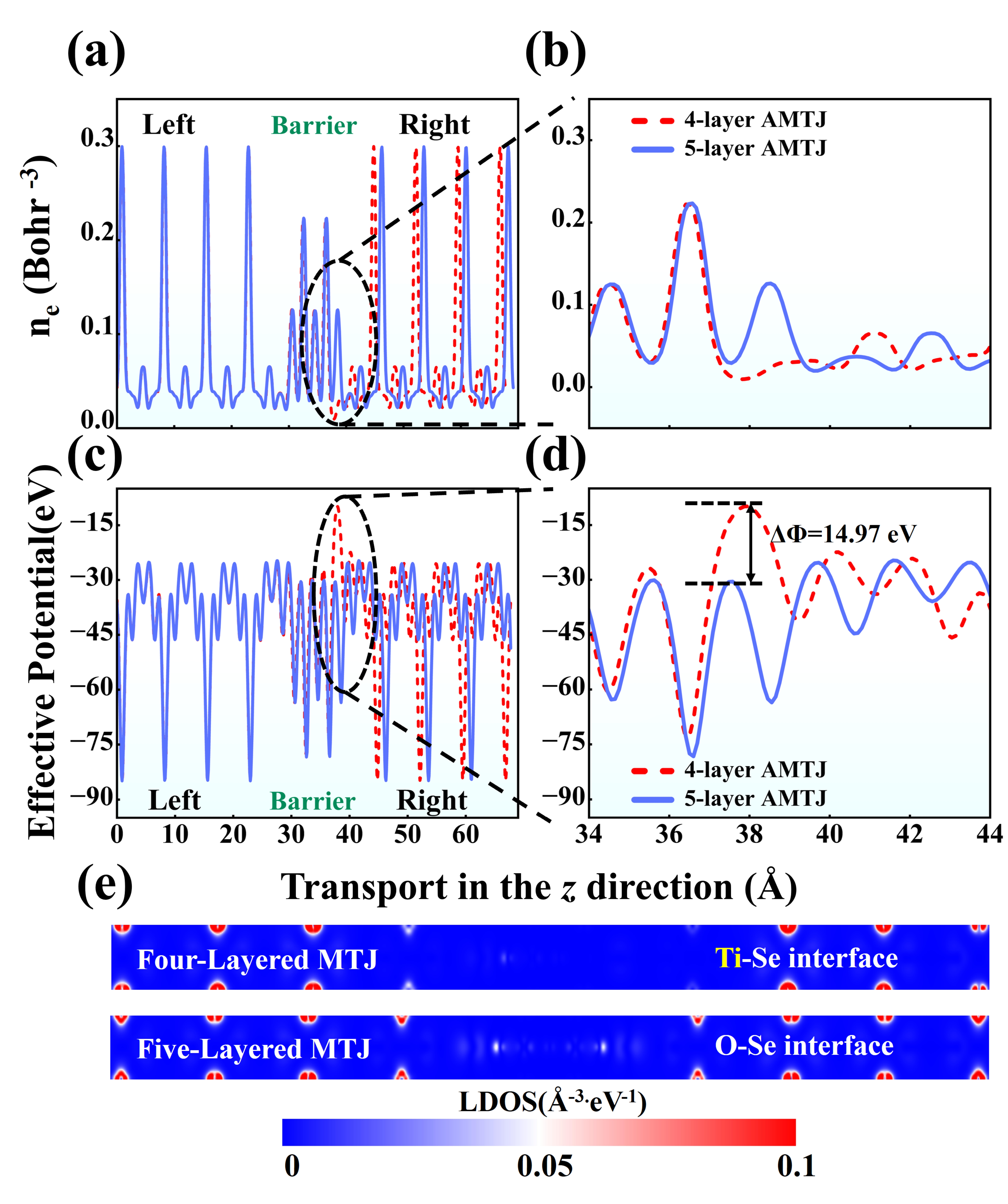}
    \caption{Electron density and effective potential of the 4-layer and 5-layer AMTJs along the transport direction. (a) Electron density $n_{e}$ along the transport direction ($z$-axis) for the two AMTJs, and (b) the corresponding magnified view of the right interface region. (c) Effective potential profiles along the transport direction, and (d) the corresponding magnified view of the right interface region. In panels (a)-(d), the red dashed lines and blue solid lines denote the 4-layer and 5-layer AMTJs, respectively. (e) Spatial distribution of the LDOS for the 4-layer (Ti–Se interface) and 5-layer (O–Se interface) AMTJs under the \textit{P} state.}
    \label{FIG-6}
\end{figure}

The transmission results intuitively demonstrate that the quantum tunneling of the $\mathrm{KV_{2}Se_{2}O/SrTiO_{3}/KV_{2}Se_{2}O}$ AMTJ is highly sensitive to the layer number\cite{Yang2025,Yang2021}. The layer dependency (odd or even) is attributed to different interfacial configurations between the electrodes and the SrTiO$_{3}$ barrier. For simplicity, the left interface of the AMTJ is fixed (taking the O-Se left interface as an example), and the right interface is dictated by the number of the layers: even-layer features a Ti-Se interfacial configuration, whereas odd-layer barrier exhibits an O-Se interface. We analyze the averaged electron density ($n_{e}$) along the transport direction, as shown in \hyperref[FIG-6]{Fig.~6(a)}. For the left electrode region, the two AMTJs are almost identical within the electrode regions. Conversely, a distinct difference is observed at the right interface of the AMTJ due to different atomic structures. \hyperref[FIG-6]{Fig.~6(b)} provides a magnified view of the right interface region. At the interface region, the $n_{e}$ of the 5-layer AMTJ (O--Se interface) is significantly higher than that of the 4-layer AMTJ (Ti--Se interface). A higher electron density indicates that more carriers could contribute to quantum transport\cite{Zeng2023,Liu2024a}, thereby increasing the probability of tunneling through the barrier\cite{Goossens2024}. 

Hence, substantial differences in effective potential between the 4-layer and 5-layer AMTJs are observed at the right interfacial region, as exhibited in \hyperref[FIG-6]{Figs.~6(c,d)}. In particular, the effective potential difference is approximately 14.97 eV. According to quantum tunneling theory, the wavefunction inside the barrier rapidly decays with respect to the effective potential\cite{Yang2021,Wang2026}. Compared with the 5-layer AMTJ, the higher potential barrier in the case of the 4-layer AMTJ makes the tunneling probability smaller. The LDOS exhibited in \hyperref[FIG-6]{Fig.~6(e)}) further support this judgment. The LDOS at the Ti–Se interface is smaller compared to that at the O–Se interface. In real space, the wavefunction within the barrier region decays more rapid in the 4-layer AMTJ\cite{Zhang2025PRB}. Consequently, the transmission coefficient in the 4-layer AMTJ is substantially lower than that in the 5-layer AMTJ. 

\begin{figure}
    \centering
    \includegraphics[width=1\linewidth]{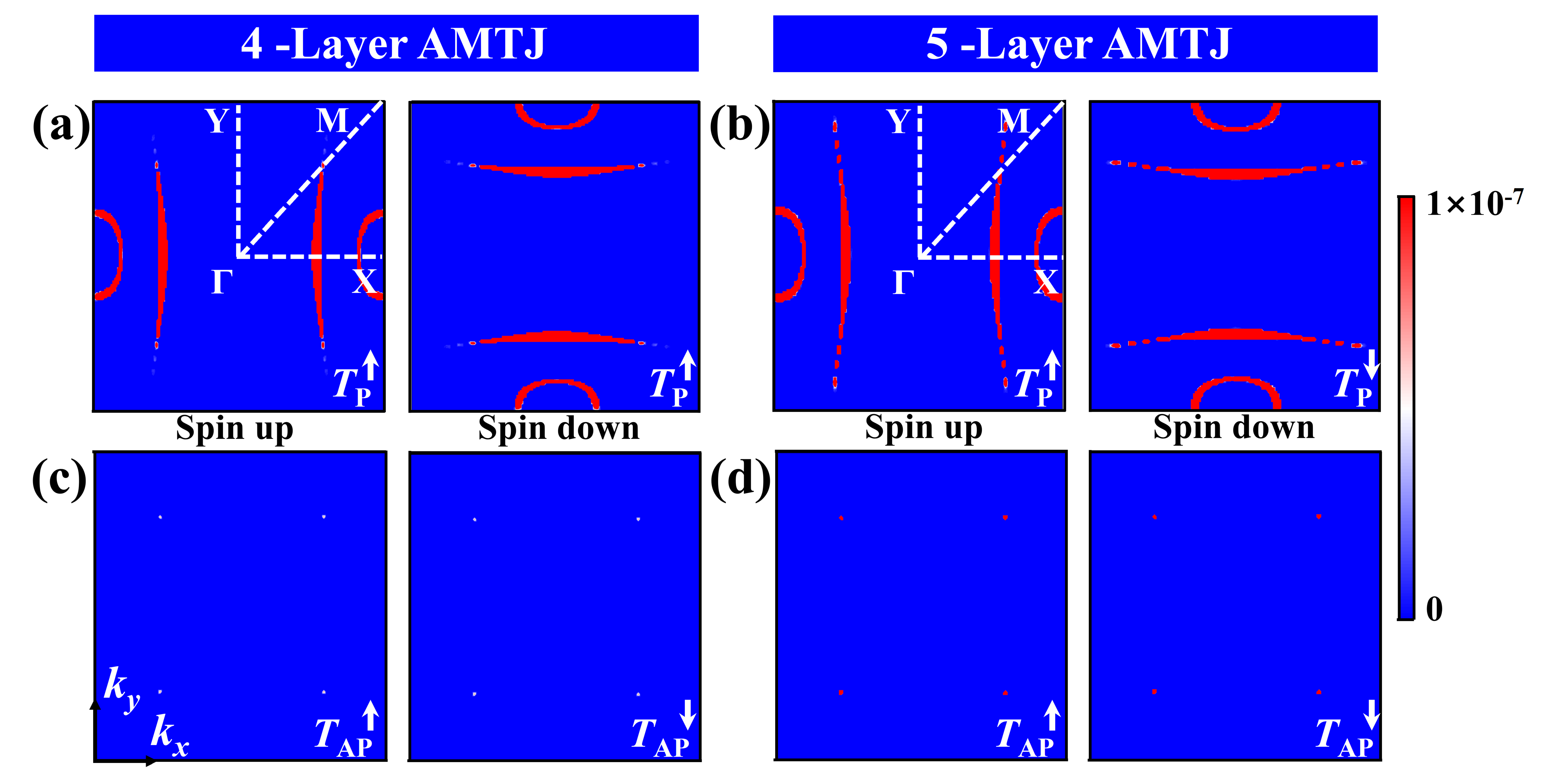}
    \caption{The $\mathbf{k}_{\parallel}$-resolved transmission spectral at the $E_{F}$. The 4-layer AMTJ under the (a) \textit{P} and (c) \textit{AP} state, respectively. The 5-layer AMTJ under the (b) \textit{P} and (d) \textit{AP} state, respectively. Red and blue areas in the 2D $\mathbf{k}_{\parallel}$ space denote high and low transmission coefficients, respectively.}
    \label{FIG-7}
    \vspace{-3mm}
\end{figure}

The $k_{\parallel}$-resolved transmission coefficients at the $E_{F}$ manifest the quantum transport in the \textit{k}-space, as presented in \hyperref[FIG-7]{Fig.~7}. Under the \textit{P} state, nonzero transmission spectra are observed for both 4-layer and 5-layer AMTJs, as shown in \hyperref[FIG-7]{Figs.~7(a,b)}. Moreover, their spin-up and spin-down transport channels are opened due to the fact that the two AM electrodes are perfectly matched under the \textit{P} state. The transport spectral are highly consistent with the projections of the anisotropic $\mathrm{KV_2Se_2O}$ Fermi surfaces shown in the \hyperref[FIG-2]{Fig.~2(c)}\cite{Shao2024}. The slight difference in transmission spectra occurs at the tails of the nodal-like arcs, leading to better transport performance in the 5-layer AMTJ. 

In the \textit{AP} state, the N\'eel vector of the two electrodes mismatches, hindering quantum transport. It is found that only four discrete nodes, i.e., the crossing points of the  four nodal-like arcs, are opened for tunneling transport\cite{Qu2025,Li2025}, as shown in \hyperref[FIG-7]{Figs.~7(c,d)}. These isolated nodes provide minimal conduction pathways, strongly suppressing the quantum transport. It is noticed that the 5-layer AMTJ exhibits enhanced transmission at these nodes when it is compared with the 4-layer AMTJ. Furthermore, the corresponding results for \textit{n}-layer AMTJs (n = 2, 3, . . . , 9) are presented in Fig. S3. These $\mathbf{k}_{\parallel}$-resolved transmission confirm the layer-dependency of the transport properties, which is consistent with \hyperref[FIG-5]{Fig.~5}. 

To evaluate the application potential of the designed AMTJs, we compare their TMR performances with recent theoretical studies and experimental results. 

\begin{figure}
    \centering
    \includegraphics[width=1\linewidth]{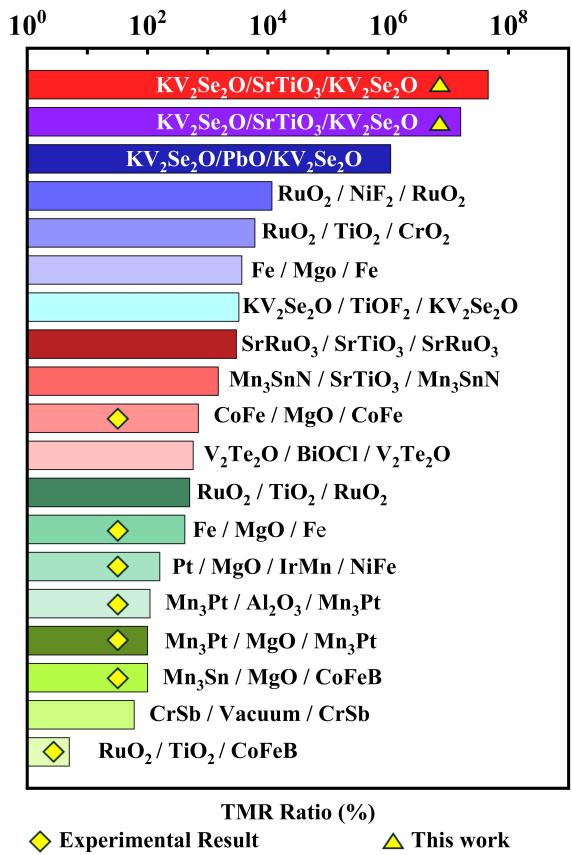}
    \caption{Benchmark of the MTJ performance. Comparison of the TMR ratios in the proposed $\mathrm{KV_2Se_2O/SrTiO_3/KV_2Se_2O}$ altermagnetic tunnel junctions (yellow triangles) with selected experimental and theoretical results from previous studies \cite{Yang2026,Zhang2026,Chi2024,Waldron2006,Samanta2024JPCM,Liu_2025,Cui2023,Shao2021,Shi2025a,Scheike2023,Scheike2021,Park2011,Shi2024,Qin2023,Chou2024a,Noh2025}. Yellow diamonds denote experimental data, and unmarked bars represent  calculated values. The horizontal axis is shown on a logarithmic scale.}
    \label{FIG-8}
    \vspace{-2mm}
\end{figure}

As shown in \hyperref[FIG-8]{Fig.~8}, our designed 4-layer $\text{KV}_2\text{Se}_2\text{O}/\text{SrTiO}_3/\text{KV}_2\text{Se}_2\text{O}$ AMTJ exhibits an enormous TMR reaching $4.6\times10^{7}$\%. In comparison, the TMR of $1.1\times10^{6}$\% is predicted to be obtained by using the $\text{KV}_2\text{Se}_2\text{O}/\text{PbO}/\text{KV}_2\text{Se}_2\text{O}$ junction\cite{Yang2026}. Moreover, the TMR value of $4.6\times10^{7}$\% prominently exceeds the theoretical limit of the conventional Fe/MgO MTJ, which is 3700\% \cite{Waldron2006}. Compared to the altermagnetic $\text{RuO}_2$ systems, where the TMR values generally range from 100$\sim$10,000\% \cite{Zhang2026,Chi2024,Shao2021}, our designed AMTJ has a superior TMR value. 

Although some emergent magnetic heterostructures are predicted to have considerable TMR, the operation temperature always limit to $10$~K in order to maintain the magnetic order. The low operation temperature practically impedes future applications for information storage\cite{Noh2025}. In contrast, the $T_{N}$ of the $\text{KV}_2\text{Se}_2\text{O}$ altermagnet electrodes is well above room temperature, ensuring stable spin transport under ambient conditions\cite{Jiang2025}. Meanwhile, the $\text{SrTiO}_3$ servicing as a NM barrier is experimentally fabricated and widely used for spintronic device\cite{Soya2025}. The $\text{KV}_2\text{Se}_2\text{O}$/$\text{SrTiO}_3$/$\text{KV}_2\text{Se}_2\text{O}$ AMTJ holds great promise for experimental realization. Our designed AMTJ possess remarkable advantages for practical applications: (1) working under room temperature; (2) ultrahigh magnetoresistance; (3) stray-field-free THz frequency\cite{Jungwirth2026}. Our study has shown that interfacial engineering is a paradigm to improve TMR performance for spintronic devices.

\section{CONCLUSIONS}
By using DFT combined with the non-equilibrium Green's function method, we systematically investigate the spin-polarized quantum transport properties of the KV$_{2}$Se$_{2}$O/SrTiO$_{3}$/KV$_{2}$Se$_{2}$O altermagnetic tunnel junctions. The transmission of the devices exhibits a pronounced even–odd oscillation, which is governed by the interfacial atomic configurations (specifically, a Ti–Se interface for even layers and an O-Se interface for odd layers). The even-layer (4-layer) AMTJ achieves a giant TMR of $4.6 \times10^{7}\%$ at the $E_{F}$. This value not only exceeds the $1.6\times10^{7}\%$ obtained in its odd-layer (5-layer) AMTJ, but also surpasses the theoretical limits of conventional magnetic tunnel junctions. These findings unveils that manipulating and optimizing the interface configuration is a strategy to tune the magnetoelectric performance of spintronic devices.

\begin{acknowledgments}
This work was supported by the National Natural Science Foundation of China (Nos. 62574115 and 62371238). The authors thank Prof. Zhiqiang Fan for helping with quantum transport calculations by ATK. We are grateful to the High-Performance Computing Centre of Nanjing University for providing the IBM Blade cluster system.
\end{acknowledgments}

\bigskip
\textbf{DATA AVAILABILITY}
\par
The data that support the findings of this article are not publicly available upon publication because it is not technically feasible and/or the cost of preparing, depositing, and hosting the data would be prohibitive within the terms of this research project. The data are available from the authors upon reasonable request.

\bigskip
\textbf{ORCID iDs}\\
Bin Xiao https://orcid.org/0009-0006-6167-8706\\
Jiawei Liu https://orcid.org/0000-0003-2071-7546\\
Hui Zeng https://orcid.org/0000-0002-7657-6714\\
Jun Zhao https://orcid.org/0000-0001-7118-4992\\

\newpage
\bibliography{prb-reference}

\end{document}